\documentclass[aps,prl,reprint,superscriptaddress]{revtex4-2}

\usepackage{graphicx}
\usepackage{amsmath}
\usepackage{amssymb}
\usepackage{caption}
\usepackage[colorlinks=true,allcolors=blue]{hyperref}

\begin{document}

\title{First Lasing and Stable Operation of a Direct-Amplification Enabled Harmonic Generation Free-Electron laser}

\author{Zheng Qi}
\altaffiliation{These authors contributed equally to this work.}
\affiliation{Shanghai Advanced Research Institute, Chinese Academy of Sciences, Shanghai 201210, China}

\author{Junhao Liu}
\altaffiliation{These authors contributed equally to this work.}
\affiliation{ShanghaiTech University, Shanghai 201210, China.}

\author{Lanpeng Ni}
\affiliation{Shanghai Institute of Applied Physics, Chinese Academy of Sciences, Shanghai 201800, China}

\author{Tao Liu}
\affiliation{Shanghai Advanced Research Institute, Chinese Academy of Sciences, Shanghai 201210, China}

\author{Zhen Wang}
\affiliation{Shanghai Advanced Research Institute, Chinese Academy of Sciences, Shanghai 201210, China}

\author{Kaiqing Zhang}
\affiliation{Shanghai Advanced Research Institute, Chinese Academy of Sciences, Shanghai 201210, China}

\author{Hanxiang Yang}
\affiliation{Shanghai Advanced Research Institute, Chinese Academy of Sciences, Shanghai 201210, China}

\author{Zhangfeng Gao}
\affiliation{Shanghai Advanced Research Institute, Chinese Academy of Sciences, Shanghai 201210, China}

\author{Nanshun Huang}
\affiliation{Zhangjiang Laboratory, Shanghai 201204, China}

\author{Si Chen}
\affiliation{Shanghai Advanced Research Institute, Chinese Academy of Sciences, Shanghai 201210, China}

\author{Hang Luo}
\affiliation{Shanghai Advanced Research Institute, Chinese Academy of Sciences, Shanghai 201210, China}

\author{Yaozong Xiao}
\affiliation{Zhangjiang Laboratory, Shanghai 201204, China}

\author{Cheng Yu}
\affiliation{Shanghai Advanced Research Institute, Chinese Academy of Sciences, Shanghai 201210, China}

\author{Yongmei Wen}
\affiliation{Shanghai Advanced Research Institute, Chinese Academy of Sciences, Shanghai 201210, China}

\author{Fei Gao}
\affiliation{Shanghai Advanced Research Institute, Chinese Academy of Sciences, Shanghai 201210, China}

\author{Yangyang Lei}
\affiliation{Shanghai Advanced Research Institute, Chinese Academy of Sciences, Shanghai 201210, China}

\author{Huan Zhao}
\affiliation{Shanghai Advanced Research Institute, Chinese Academy of Sciences, Shanghai 201210, China}

\author{Yanyan Zhu}
\affiliation{Shanghai Advanced Research Institute, Chinese Academy of Sciences, Shanghai 201210, China}

\author{Liping Sun}
\affiliation{Shanghai Advanced Research Institute, Chinese Academy of Sciences, Shanghai 201210, China}

\author{Weiyi Yin}
\affiliation{Shanghai Advanced Research Institute, Chinese Academy of Sciences, Shanghai 201210, China}

\author{Xingtao Wang}
\affiliation{Shanghai Advanced Research Institute, Chinese Academy of Sciences, Shanghai 201210, China}

\author{Taihe Lan}
\affiliation{Shanghai Advanced Research Institute, Chinese Academy of Sciences, Shanghai 201210, China}

\author{Xiaoqing Liu}
\affiliation{Shanghai Advanced Research Institute, Chinese Academy of Sciences, Shanghai 201210, China}

\author{Lie Feng}
\affiliation{Shanghai Advanced Research Institute, Chinese Academy of Sciences, Shanghai 201210, China}

\author{Wenyan Zhang}
\affiliation{Shanghai Advanced Research Institute, Chinese Academy of Sciences, Shanghai 201210, China}

\author{Ximing Zhang}
\affiliation{Shanghai Advanced Research Institute, Chinese Academy of Sciences, Shanghai 201210, China}

\author{Bin Li}
\affiliation{Shanghai Advanced Research Institute, Chinese Academy of Sciences, Shanghai 201210, China}

\author{Chao Feng}
\email{fengc@sari.ac.cn}
\affiliation{Shanghai Advanced Research Institute, Chinese Academy of Sciences, Shanghai 201210, China}

\author{Bo Liu}
\affiliation{Shanghai Advanced Research Institute, Chinese Academy of Sciences, Shanghai 201210, China}

\author{Zhentang Zhao}
\email{zhaozt@sari.ac.cn}
\affiliation{Shanghai Advanced Research Institute, Chinese Academy of Sciences, Shanghai 201210, China}

\begin{abstract}
Seeded free-electron lasers (FELs) capable of operating at repetition rates up to the MHz level are in high demand for advanced time-resolved spectroscopies, which require both full longitudinal coherence and high average photon flux in the extreme ultraviolet (EUV) and x-ray regimes. However, conventional external-seed laser systems cannot sustain MHz operation with sufficient hundreds of megawatts peak power requirement due to their limited total power. Here, we report the first lasing and stable operation of a direct-amplification-enabled harmonic generation FEL driven by a weak seed laser with MW-level peak power. Beginning with an ultraviolet seed laser with only 0.75 $\mu J$ pulse energy, we demonstrate its direct amplification to over 10 $\mu J$ within an 8-meter-long modulator. We observe coherent harmonic generation up to the $12th$ harmonic of the seed and achieve saturation of the $7th$ harmonic in the radiator. These results represent a crucial milestone toward the realization of MHz-class, fully coherent EUV and x-ray light sources.
\end{abstract}

\maketitle

X-ray free-electron lasers (XFELs) have been revolutionary tools and open a new era in the study of nature fundamentals in an unprecedented spatio-temporal resolution, due to their advantage in producing ultrashort pulse duration and ultrahigh peak brightness x-ray pulses~\cite{pellegrini2016physics,bostedt2016linac,feng2018review,huang2021features}. Over the past two decades, rapid development of XFEL facilities worldwide has enabled groundbreaking experiments in areas ranging from time-resolved spectroscopy and coherent diffraction imaging to nanoscale microscopy and quantum materials research. Today, eight XFEL user facilities are in operation, including five hard x-ray FEL facilities~\cite{emma2010first,ishikawa2012compact,kang2017hard,prat2020compact,decking2020mhz} and three soft x-ray FEL facilities~\cite{ackermann2007operation,allaria2012highly,liu2021sxfel}, and three more are under construction~\cite{zhu2017sclf,galayda2018lcls,wang2023physical}. 

Most XFEL facilities work on the basis of the self-amplified spontaneous emission (SASE) mechanism~\cite{kondratenko1980,bonifacio1984}, in which shot-noise in the electron beam seeds exponential growth of the radiation field. While SASE can achieve extremely high pulse energies, its stochastic nature leads to limited longitudinal coherence and significant pulse-to-pulse fluctuations. Various approaches, self-seeding~\cite{FELDHAUS1997341,geloni2011novel,liu2023cascaded}, mode-locking~\cite{thompson2008mode,PhysRevLett.133.205001} and, in particular, external seeding, have been developed to improve the longitudinal coherence. In seeded schemes such as high-gain harmonic generation (HGHG)~\cite{yu1991generation,yu2000high,allaria2012highly} and echo-enabled harmonic generation (EEHG)~\cite{xiang2009echo,zhao2012first,rebernik2019coherent}, external lasers imprint precise energy modulations on the electron beam, which is then converted via dispersion sections into a high-contrast bunching pattern and used to generate coherent pulses at harmonics of the seed laser. Inheriting the good properties from the external seed laser, seeded FELs can deliver stable fully coherent FEL pulses within the extreme ultraviolet (EUV) and x-ray regime.

State-of-the-art experiments increasingly demand fully coherent EUV and x-ray FEL pulses at MHz repetition rates, for example, to perform advanced time-resolved spectroscopy~\cite{macklin1996imaging,instruments3030047}, coherent diffraction imaging~\cite{Pedersoli_2013,Capotondi:ig5025,helfenstein2017coherent}, EUV microscopy~\cite{bencivenga2015four}, and studies of ultrafast dynamics in nanostructure and nanodevice~\cite{mochi2019lensless,kudilatt2020quantum}. Superconducting linacs today routinely deliver electron beams at MHz repetition rates, and MHz-class SASE-FELs have already been demonstrated~\cite{ackermann2007operation,decking2020mhz}. However, scaling seeded FELs to these rates remains challenging: generating sufficient energy modulation normally requires seed-laser peak powers of hundreds of megawatts, which restricts repetition rates to the kilohertz regime.

\begin{figure*}[htbp]
    \centering
    \includegraphics[width=0.9\textwidth]{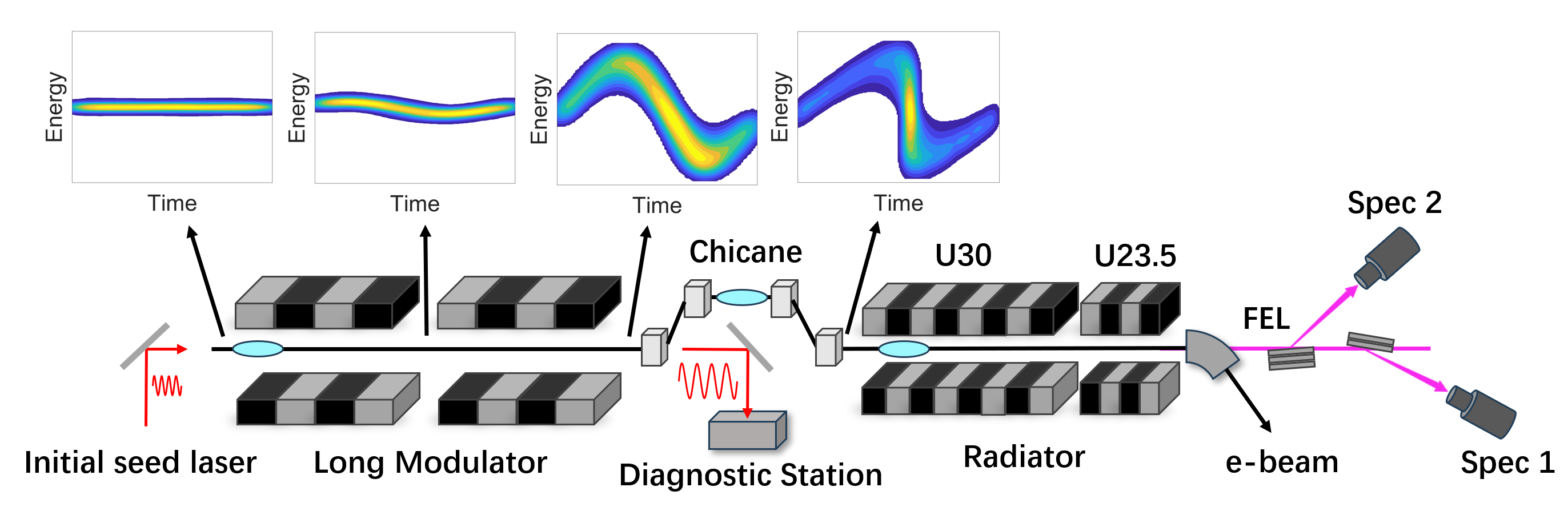}  
    \caption{Diagram of the experimental setup and the phase space evolution of the electron beam}
    \label{fig:setup}
\end{figure*}

Several approaches have been proposed to significantly reduce the required power of the seed laser and  thereby enable the MHz repetition rate operation of seeded FELs, such as the optical resonator~\cite{reinsch2012radiator,li2018high,petrillo2020coherent,mirian2021high}, the angular dispersion enhanced microbunching~\cite{feng2017storage,wang2019angular},  the optical-klystron HGHG~\cite{paraskaki2021high} and the self-modulation scheme~\cite{yan2021self,yang2023high}. To date, only the self‐modulation principle has been demonstrated experimentally, while the other methods remain to be validated. However, the self-modulation scheme requires an additional modulator and a large dispersion section to shape the electron beam, which can exacerbate microbunching instabilities~\cite{SALDIN2004355, PhysRevAccelBeams.20.120701} and render the emitted pulse coherence more susceptible to electron beam imperfections~\cite{PhysRevAccelBeams.20.060702,chirponseededFELs}.

In this Letter, we experimentally demonstrate the first lasing and stable operation of a direct-amplification enabled harmonic generation (DEHG)~\cite{wang2022high} FEL at the Shanghai soft x-ray free-electron laser facility (SXFEL)~\cite{liu2021sxfel}, by using a significantly reduced power of the seed laser. The experimental setup and the simulated phase space evolution of the electron beam with Genesis 1.3~\cite{reiche1999genesis} are shown in Fig.~\ref{fig:setup}. The DEHG configuration comprises a long modulator, a compact dispersion chicane, and a downstream radiator. A weak ultraviolet seed laser initially co-propagates with the electron bunch through the modulator, where the seed is amplified and imprints a strong energy modulation onto the beam. As a result of the high-gain interaction, the mean energy of the electron beam shifts slightly downward, visible in the phase-space distributions. In the dispersion section, this energy modulation is converted into a high-contrast density modulation, up-converting the radiation to higher harmonics. Finally, the tailored electron microbunches emit and coherently amplify radiation at the chosen harmonic within the radiator. Compared to other high-repetition-rate seeding techniques, DEHG offers a simpler beamline design, preserves longitudinal coherence of the seed (since no additional dispersion sections are introduced) and in principle could reduce the seed laser power requirement to just above the shot-noise threshold with a sufficient long modulator. 

The main parameters used in the experiment are summarized in Tab.~\ref{tab:Main Parameters}. The electron beam energy was tuned to 605 $MeV$ with a transverse emittance of about 1.5 $mm \cdot mrad$ and a peak current exceeding 600 $A$. An ultraviolet laser pulse with central wavelength of 266.9 $nm$ (the third harmonic of the Ti: Sapphire laser) and an initial pulse duration of about 200 $fs$ was employed in the experiment. To preserve the spatio-temporal quality of the laser beam through a 30-meter-long optical transport line, the pulse was stretched to 5 ps in the laser laboratory and then recompressed immediately before injection as the seed laser. The seed laser pulse energy can be tuned and measured online during the experiment. The long modulator comprised two 4 m undulator segments with the period of 68 $mm$ separated by focusing magnets and a phase shifter. A dispersion section (chicane) with tunable dispersion strength $R_{56}$ from 0 to 2 $mm$ converted the energy modulation into density modulation. At the midpoint of the chicane, the amplified seed pulse was extracted onto a diagnostic station, equipped with a fluorescent screen, an energy detector, and a fiber spectrometer, where its transverse profile, pulse energy, and spectrum were recorded. High-harmonic radiation was generated in a main radiator of five 3-meter-long undulator segments with 30 $mm$ period (U30), followed by downstream 23.5 $mm$-period undulator segments (U23.5) for cascading to shorter wavelengths via harmonic lasing~\cite{photonics8020044}. The FEL properties were characterized downstream by an insertion screen between U30 and U23.5 (for beam profile), a calibrated photodiode (for pulse energy), a commercial EUV to soft-x-ray spectrometer (Spec1, broad coverage, moderate resolution), and a custom high-resolution spectrometer (Spec 2) optimized for 3-17 $nm$ in the x-ray beamline.
\begin{table}[htbp] 
\caption{Main parameters used in the experiment.}
\label{tab:Main Parameters}
\begin{ruledtabular}
\begin{tabular}{cc}
Electron beam \\
Energy & 605 MeV \\
Charge & 500 pC \\
Bunch length (FWHM) & 500 fs \\
Peak current & $>$600 A \\
Project emittance & 1.5 mm$\cdot$mrad \\
\hline
Seed laser \\
Wavelength & 266.9 nm \\
\hline
Long Modulator \\
$N_s \times N_p \times \lambda_u$ & 2 $\times$ 58 $\times$ 68 mm \\
\hline
Radiator \\
$N_s \times N_p \times \lambda_u$ for U30 & 5 $\times$ 100 $\times$ 30 mm \\
$N_p \times \lambda_u$ for U23.5 & 120 $\times$ 23.5 mm \\
\end{tabular}
\end{ruledtabular}
\end{table}

\begin{figure*}[htbp]
  \centering
  \begin{minipage}[t]{0.3\textwidth}
    \includegraphics[width=\textwidth]{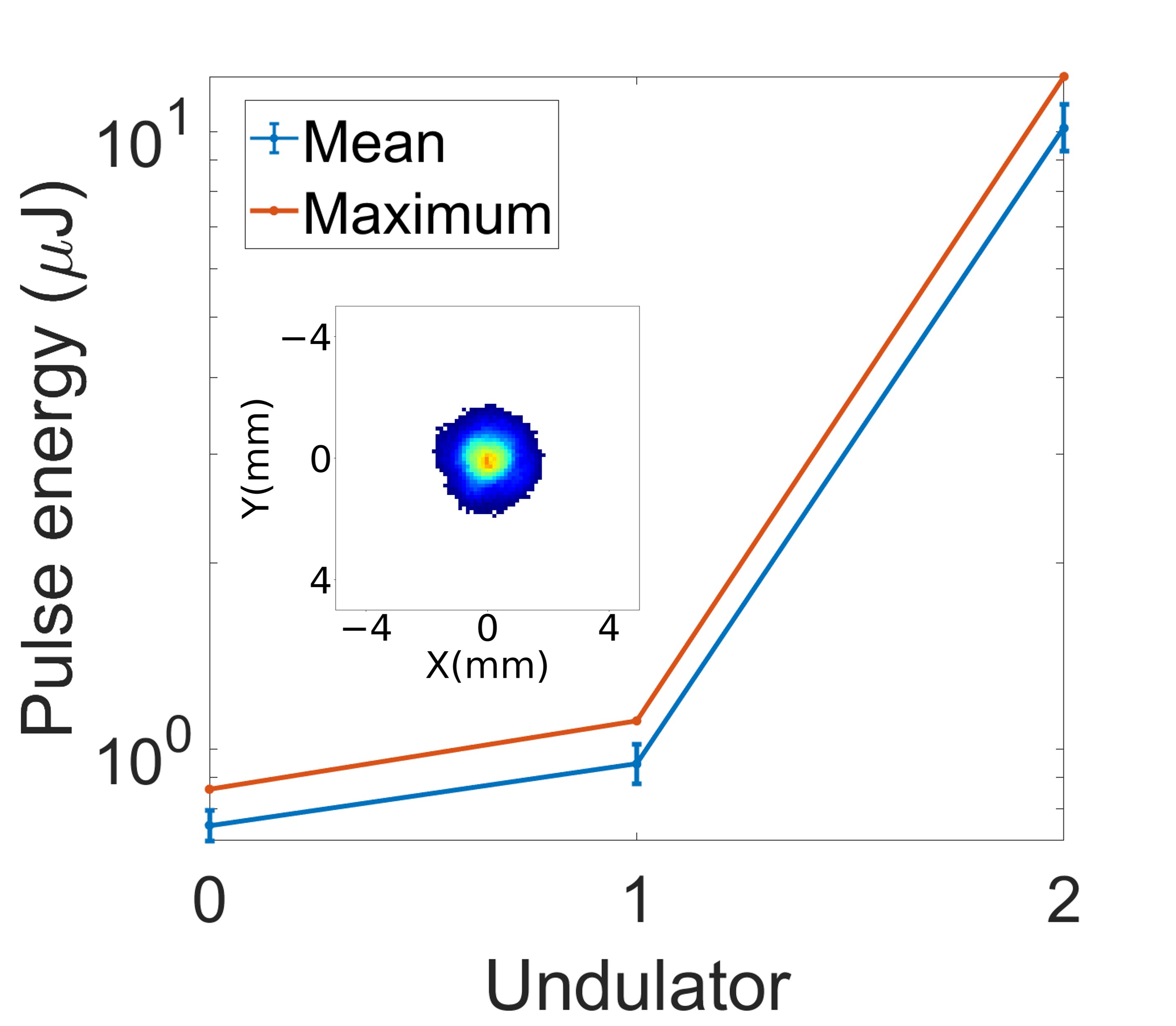}
    \caption*{(a)}
  \end{minipage}
  \hspace{0.0001\textwidth}  
  \begin{minipage}[t]{0.32\textwidth}
    \includegraphics[width=\textwidth]{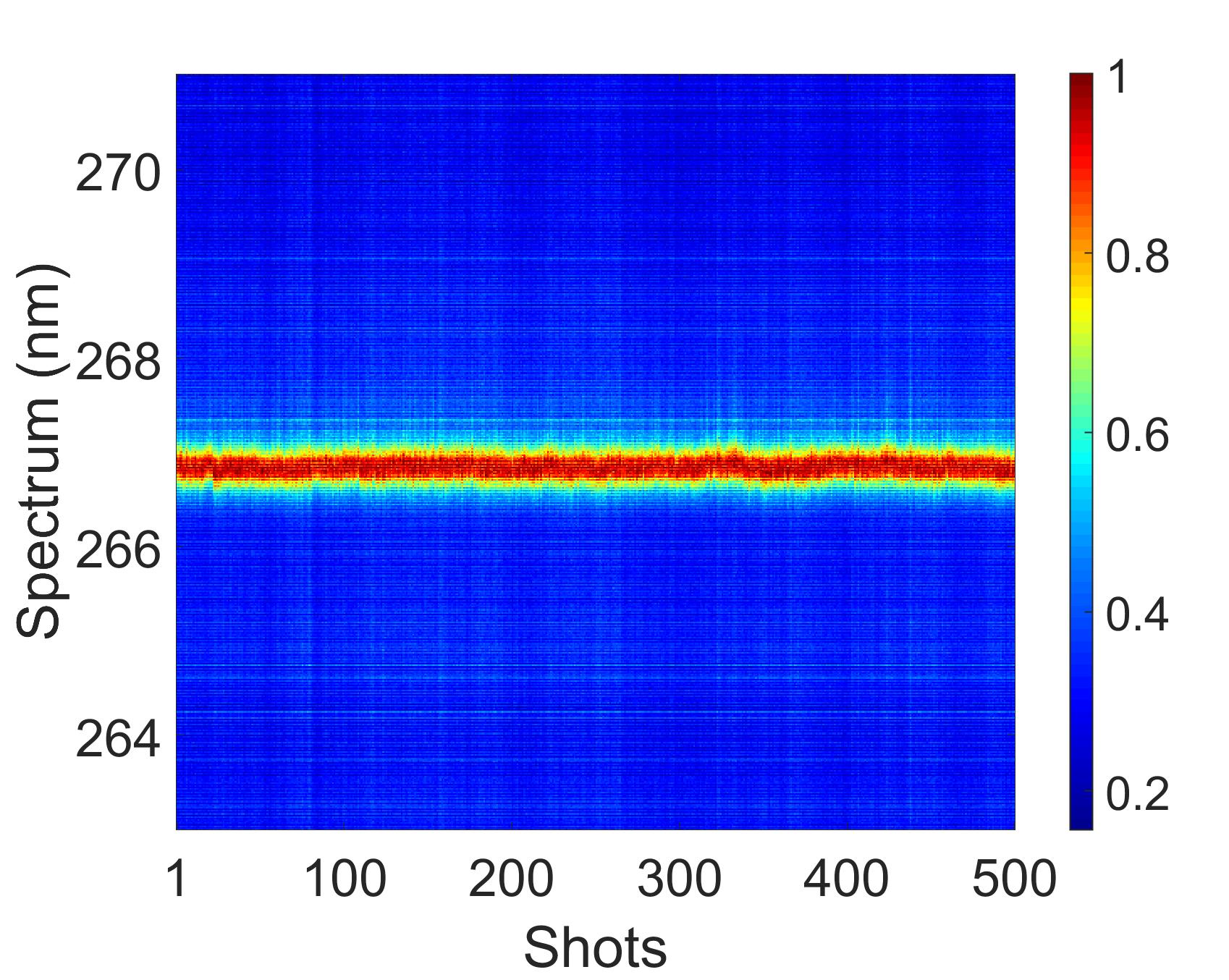}
    \caption*{(b)}
  \end{minipage}
  \hspace{0.0001\textwidth}
  \begin{minipage}[t]{0.32\textwidth}
    \includegraphics[width=\textwidth]{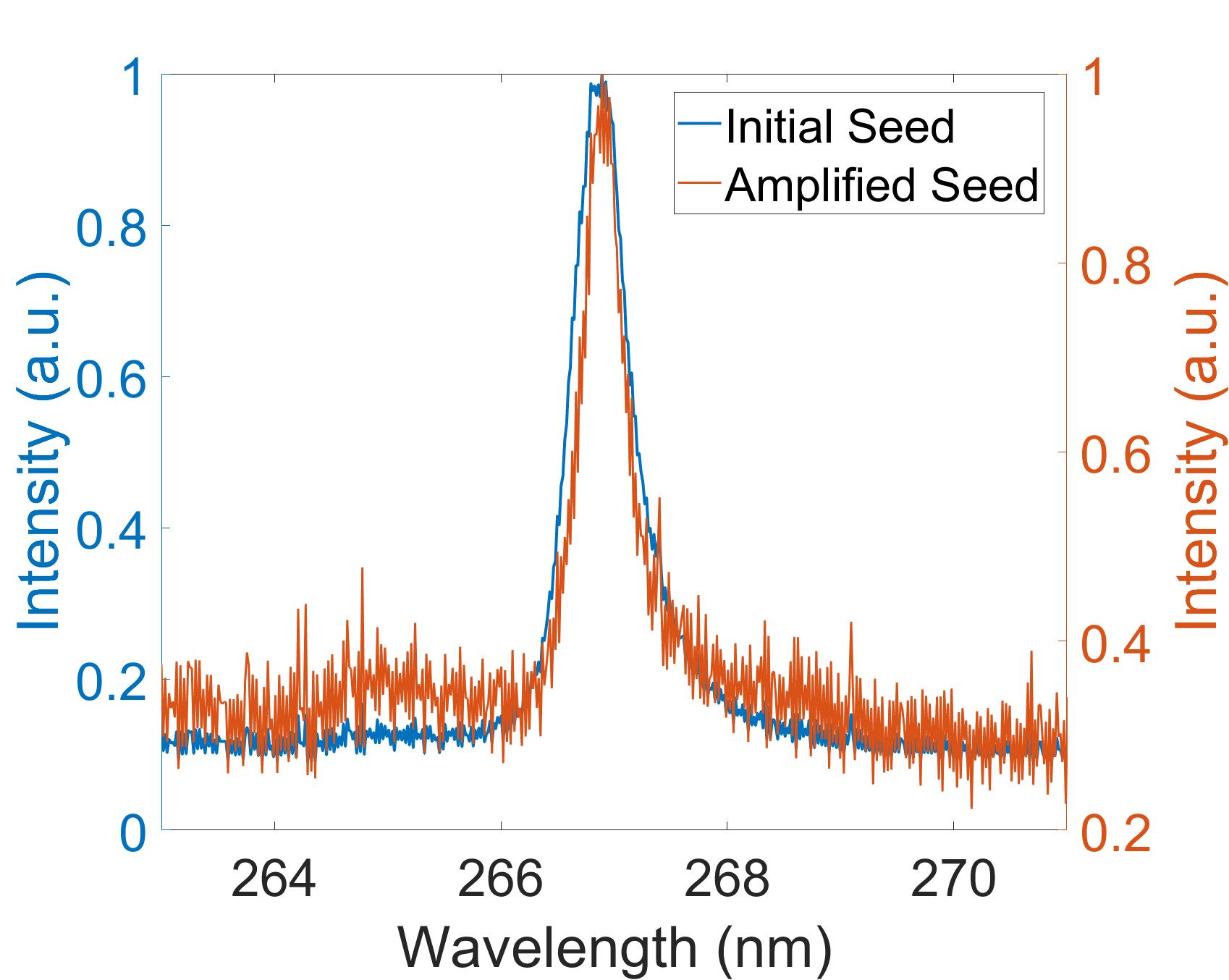}
    \caption*{(c)}
  \end{minipage}
  \caption{The measured amplified seed laser results after the long modulator section. (a) Gain curve and the transverse spot. (b) Spectra of 500 consecutive shots. (c) Spectra comparing to the initial seed.}
  \label{fig:Amplified Seed}
\end{figure*}
The amplification results of the initial seed laser in the long modulator are shown in Fig.~\ref{fig:Amplified Seed}. With both undulators resonated at 266.9 $nm$, the seed laser pulse energy was amplified from 0.75 $\mu J$ to over 10 $\mu J$ in average, as shown in Fig.~\ref{fig:Amplified Seed}(a). And the maximum 12.3 $\mu J$ is corresponding to a magnification factor of $16.4$. The shot-to-shot fluctuation of the amplified seed laser power is $7.1\%$ (rms), indicating that the amplification process is quite stable. The peak power of the initial seed laser is only several megawatts, well below the 100 $MW$ level typically required for a seeded FEL. The amplified beam exhibited an ideal Gaussian transverse profile (inset, Fig.~\ref{fig:Amplified Seed}(a)). Spectral measurements over 500 shots, as shown in  Fig.~\ref{fig:Amplified Seed}(b), confirmed that the amplified seed spectra remained highly stable after amplification. The bandwidth of the amplified seed was unchanged from its initial value, approximately $2.1 \times 10^{-3}$ (9.8 $meV$), as shown in Fig.~\ref{fig:Amplified Seed}(c), demonstrating perfect preservation of the temporal coherence of the seed laser. Moreover, all measured spatio-temporal properties of the initial seed were inherited by the amplified pulse, confirming its suitability for reuse, for example, as the second seed in an EEHG setup~\cite{wang2022high}.

\begin{figure}[htbp]
  \centering
  \begin{minipage}[t]{0.235\textwidth}
    \includegraphics[width=\textwidth]{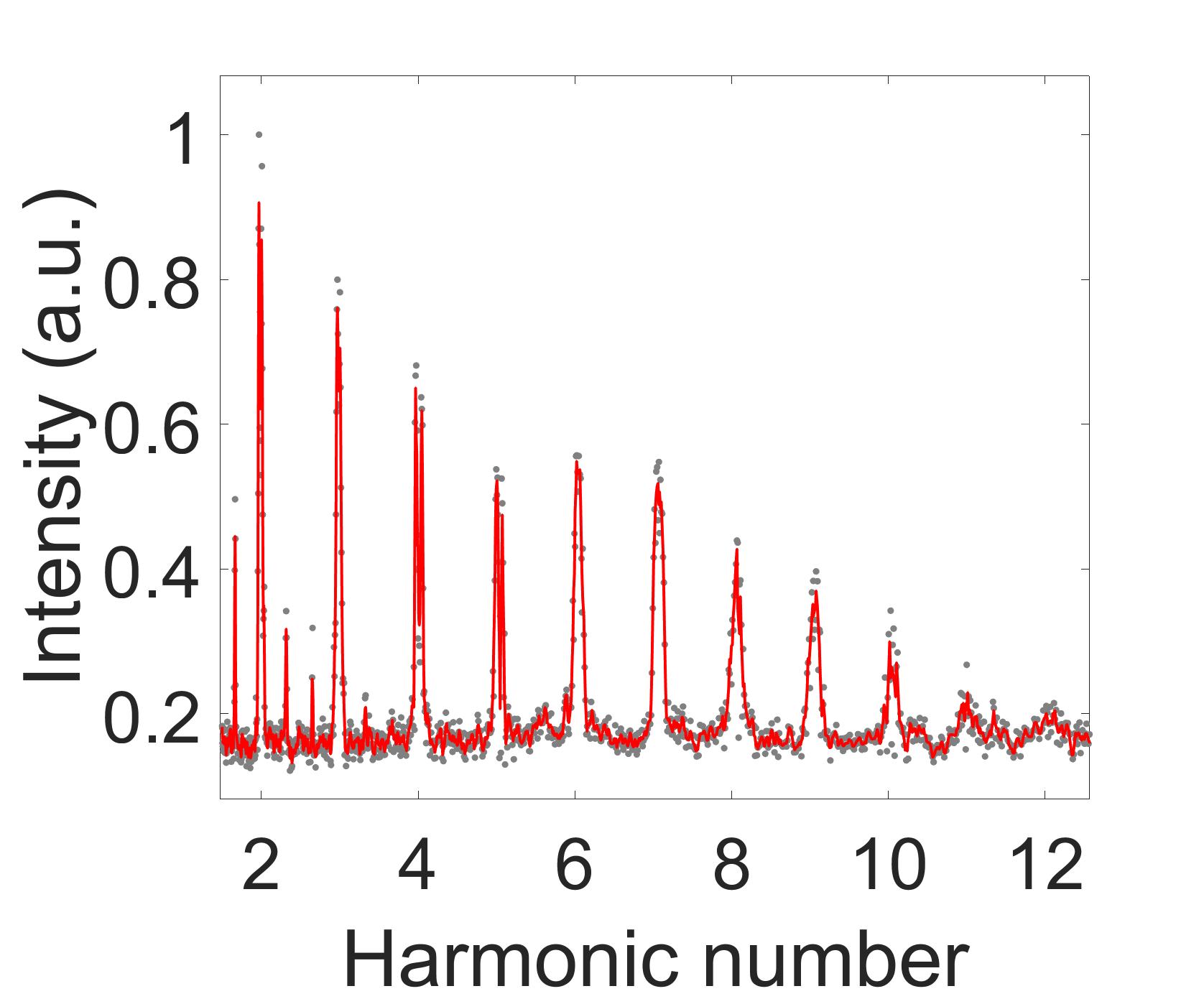}
    \caption*{(a)}
  \end{minipage}
  \hspace{0.0001\textwidth}  
  \begin{minipage}[t]{0.235\textwidth}
    \includegraphics[width=\textwidth]{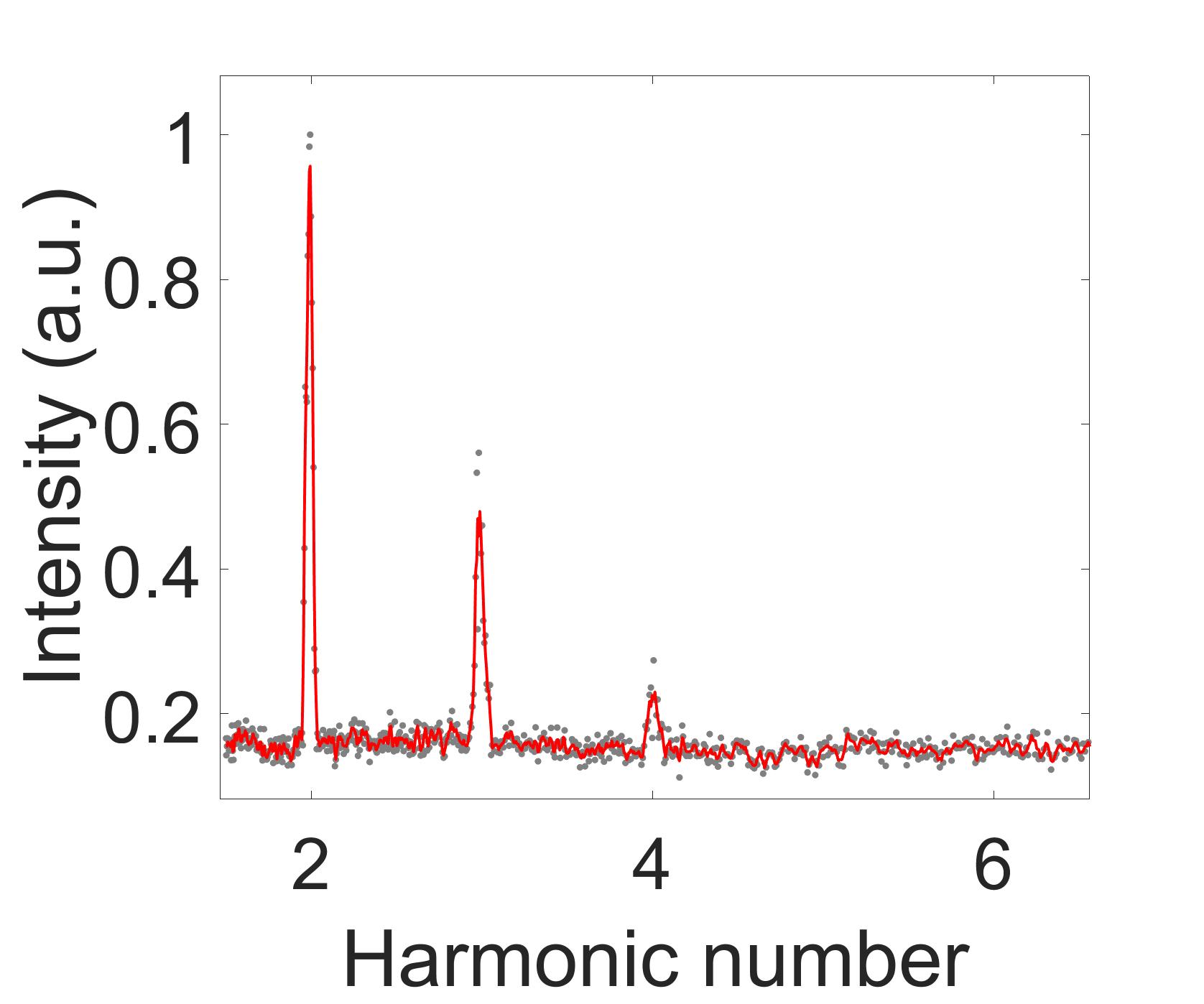}
    \caption*{(b)}
  \end{minipage}
    \caption{Comparison of coherent radiation intensities at various harmonic numbers for DEHG (a) and HGHG (b). DEHG employs two U68 modulator sections while HGHG only uses the first one. Points are measured intensities by a downstream photodiode and red curves are the fitting results.}
  \label{fig:Harmonic}
\end{figure}
Along with the amplification of the seed laser, significant energy modulation was imposed on the electron beam. Microbunched electron beam will be formed in the dispersion section and coherent high-harmonic radiation was subsequently emitted in the following radiator. The first radiator undulator employed a dual-period configuration, with switchable magnetic arrays of 50 $mm$ and 30 $mm$ period lengths, to enable scanning and optimization of harmonic radiation intensities from low harmonics. As shown in Fig.~\ref{fig:Harmonic}(a), coherent signals were observed up to the $12th$ harmonic with the optimized dispersion strength $R_{56}$ = 36.4 $\mu m$. The slice energy spread was measured at 80–100 $keV$ based on the coherent harmonic generation method~\cite{PhysRevSTAB.14.090701}. The inferred energy‐modulation amplitude within the long modulator was approximately 7, a level sufficient to drive both harmonic generation and FEL lasing. For comparison, the gap of the second undulator was detuned, limiting amplification to the first undulator segment, and thus emulating a conventional HGHG configuration. In this case, the dispersion strength $R_{56}$ was increased to 311 $\mu m$ to optimize the $4th$ harmonic bunching, implying an energy-modulation amplitude of only about 1, which is consistent with simulations and the measurement results in Fig.~\ref{fig:Amplified Seed}(a). As shown in Fig.~\ref{fig:Harmonic}(b), coherent radiation was then observed only up to the $4th$ harmonic.      

\begin{figure}[htbp]
  \centering
  \begin{minipage}[t]{0.234\textwidth}
    \includegraphics[width=\textwidth]{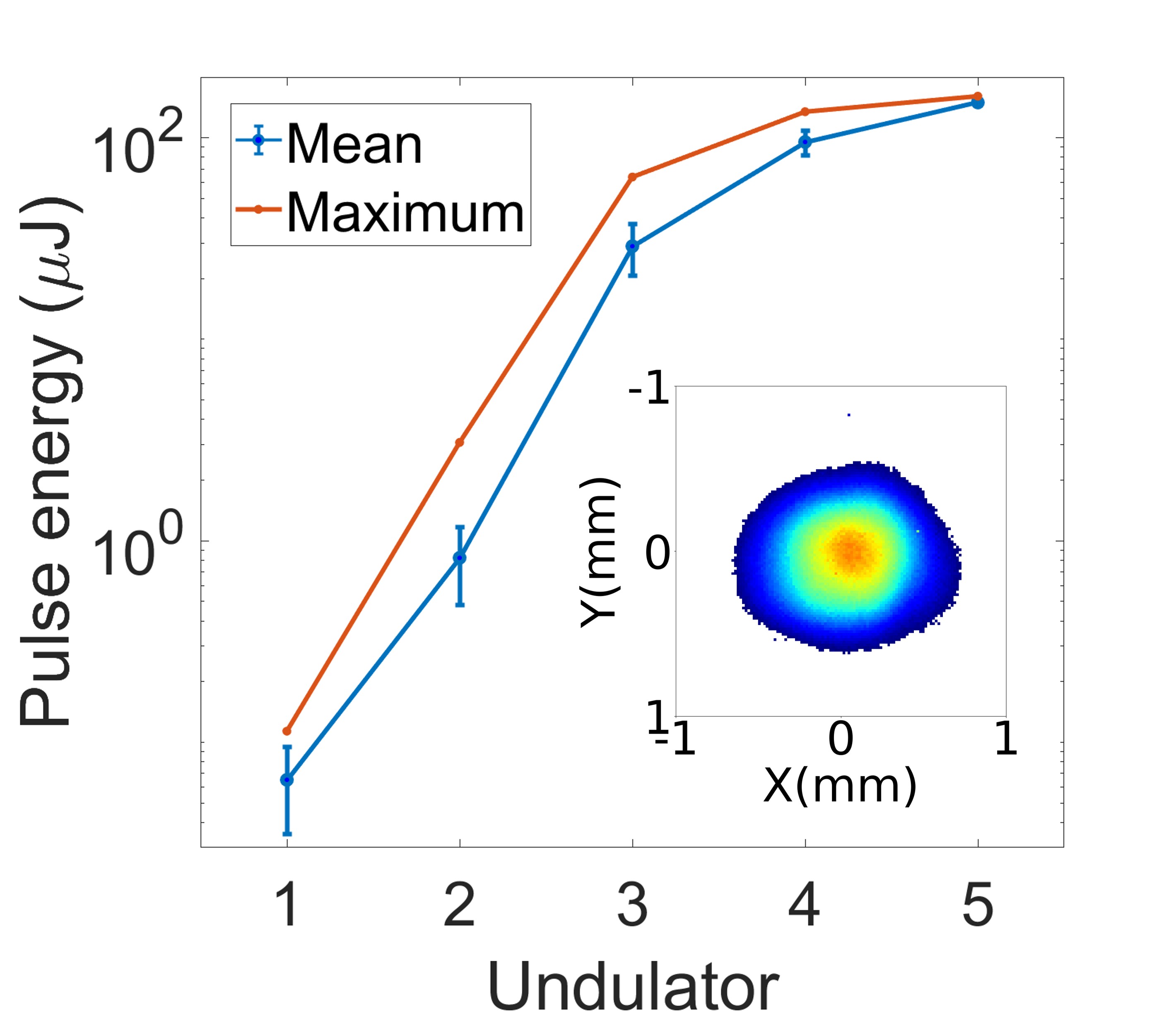}
    \caption*{(a)}
  \end{minipage}
  \hspace{0.0001\textwidth}  
  \begin{minipage}[t]{0.234\textwidth}
    \includegraphics[width=\textwidth]{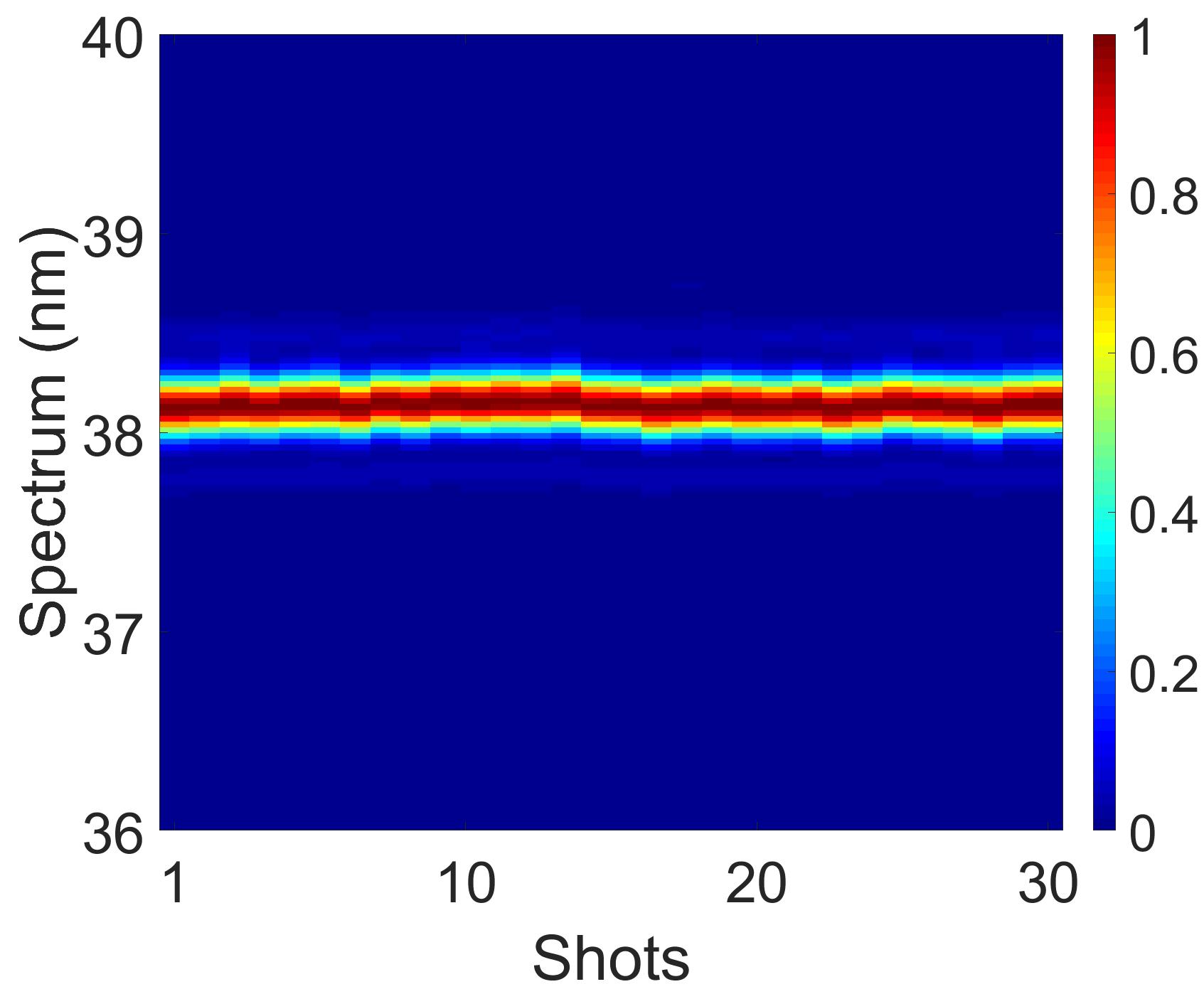}
    \caption*{(b)}
  \end{minipage}
    \caption{Performance of the DEHG FEL lasing at the 7th harmonic of the seed laser. (a) The FEL gain curve and the transverse spot. (b) The spectra of 30 consecutive shots.}
  \label{fig:DEHG-7th}
\end{figure}
To verify the principle and demonstrate the performance of the DEHG FEL, the remaining U30 segments were used to amplify the $7th$ harmonic. The resulting gain curve, transverse spot and spectra from 30 consecutive shots are illustrated in Fig.~\ref{fig:DEHG-7th}. The pulse energy is saturated within the $5th$ undulator, with a maximal saturation energy of approximately 160 $\mu J$ and a shot-to-shot stability of $5.5\%$ (rms). The insertion shows the transverse profile of the FEL beam that is ideal Gaussian with a spot size about 0.6 $mm$ (FWHM). Spectral measurements performed based on the commercial spectrometer reveal a bandwidth of 5.9 $\times 10^{-3}$ for the 38.1 $nm$ radiation.         

\begin{figure}[htbp]
  \centering
  \begin{minipage}[t]{0.235\textwidth}
    \includegraphics[width=\textwidth]{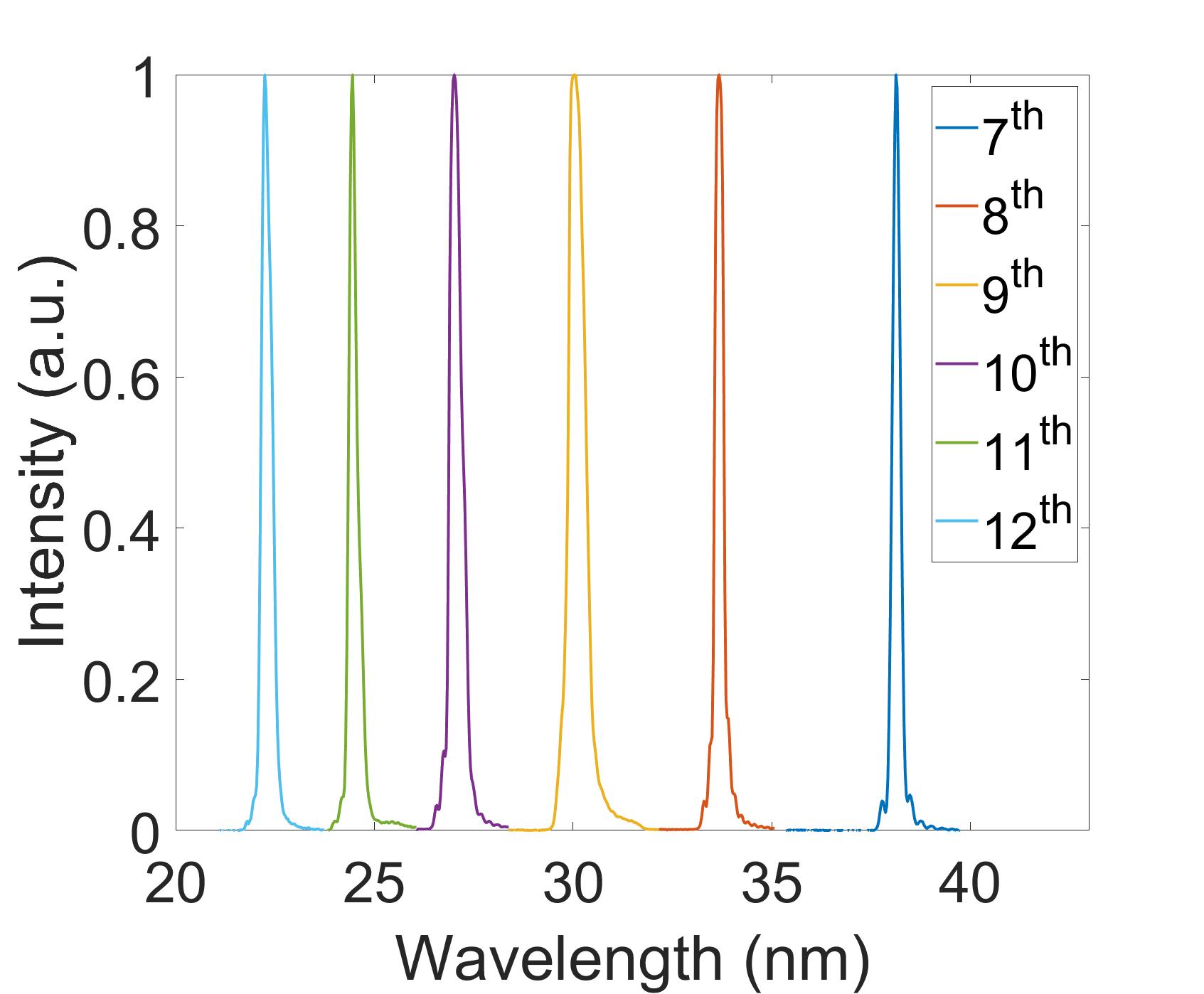}
    \caption*{(a)}
  \end{minipage}
  \hspace{0.0001\textwidth}  
  \begin{minipage}[t]{0.235\textwidth}
    \includegraphics[width=\textwidth]{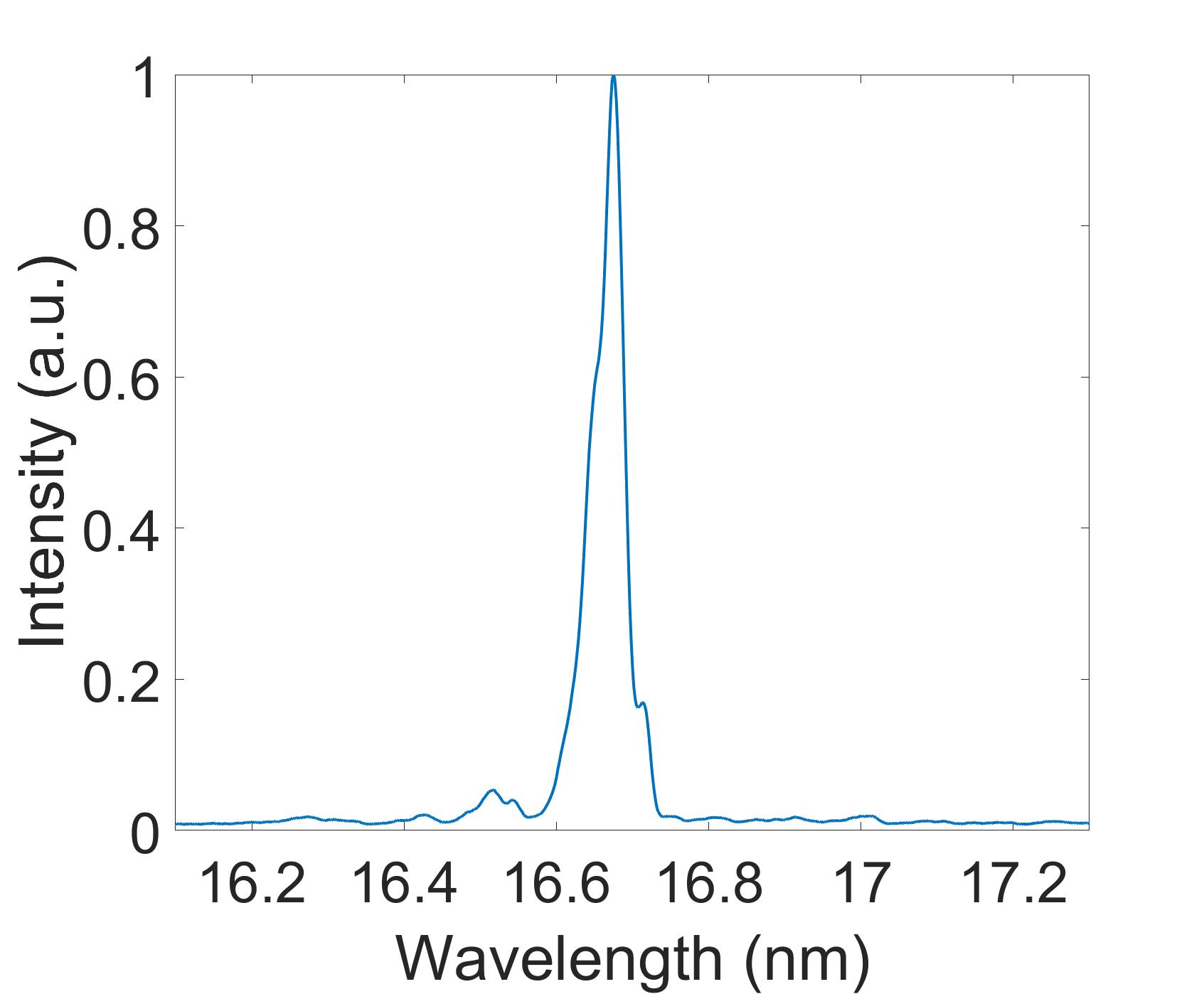}
    \caption*{(b)}
  \end{minipage}
    \caption{(a) Measured DEHG FEL spectra from 7th to 12th harmonics. (b) Measured single-shot spectrum of the 16th harmonic radiation.}
  \label{fig:7-12thSpectra}
\end{figure}
 By continuously tuning the magnetic gap of the radiator, spectra from 7th to 12th harmonics were recorded, as summarized in Fig.~\ref{fig:7-12thSpectra}(a). The measured bandwidths for these harmonics were comparable, limited by the 0.1 $nm$ resolution of the spectrometer. To achieve higher spectral resolution, we need to use the custom online x-ray spectrometer~\cite{Yang:rv5181} optimized for the 3–17 $nm$ range. By maintaining $8th$-harmonic microbunching in the U30 segments and employing a downstream U23.5 segment resonant at twice the U30 fundamental, a coherent $16th$-harmonic signal at 16.68 nm was produced, just within the detectable range of the spectrometer. The high-resolution spectrum, shown in Fig.~\ref{fig:7-12thSpectra}(b), exhibits a single, well-defined peak with a bandwidth of 6.5 $\times 10^{-4}$ (48 $meV$), approximately five times broader than the seed. This broadening is primarily attributed to pulse shortening effect for high harmonics~\cite{Finetti.2017b} and residual frequency chirp in the seed~\cite{PhysRevSTAB.15.030702}. In seeded FEL operations, the output pulse duration scales with the high-harmonic bunching factor, ranging between $1/\sqrt{n}$ and $7/(6n^{1/3})$ relative to the seed, depending on the extent of FEL amplification~\cite{Finetti.2017b}. For the $16th$ harmonic, a fourfold reduction in pulse duration is anticipated, considering that the radiation is far from saturation. Under these assumptions, the time-bandwidth product of the harmonic radiation pulse closely matches that of the seed, indicating well-preserved longitudinal coherence in the DEHG-FEL.

In summary, the DEHG-FEL has been successfully demonstrated as a robust external-seeding technique that combines substantial seed amplification with high-order harmonic generation. The experimental results, featuring more than an order-of-magnitude increase in seed pulse energy, highly stable spectra, and coherent lasing up to the $12th$ harmonic, confirm the viability of DEHG as a low-power seeding strategy. It is noteworthy that the seed laser amplification factor can be further enhanced by optimizing modulation segment parameters and utilizing higher-quality electron beams. Simulation studies predict that such optimizations could reduce external seed-power requirements by three orders of magnitude~\cite{wang2022high}. Moreover, the amplified seed pulse inherits the full coherence of the original laser, making it reusable, for instance, as the second seed in an EEHG configuration. This approach paves the way for fully coherent soft-x-ray FELs operating at MHz-class repetition rates using commercially available laser systems, thereby broadening access to ultrafast, high-resolution experiments across a variety of scientific disciplines. 

The capacity of DEHG to amplify very weak seeds makes it particularly suitable for seeding with short-wavelength and low-power lasers, such as nJ-level high-harmonic generation pulses from noble gases~\cite{Labaye:17, dunning2011design}. Such capabilities are expected to extend the applicability of externally seeded FELs to new wavelength domains, unlocking advanced spectroscopic and scattering techniques across chemistry, biology, and materials science.

\begin{acknowledgments}
The authors would like to thank D. Xiang, E. Allaria, X. Wang, and B. Faatz for helpful discussions. This work was supported by the SXFEL facility, the National Key Research and Development Program of China (Grant No.2024YFA1612104), National Natural Science Foundation of China (Grant No.12435011, 12405363), Project for Young Scientists in Basic Research of Chinese Academy of Sciences (YSBR-115, YSBR-091), and Shanghai Municipal Science and Technology Major Project.
\end{acknowledgments}

\bibliographystyle{apsrev4-2}
\bibliography{refs}  

\end{document}